\documentclass[12pt,preprint]{aastex6}
\usepackage{graphicx}

\begin{document}  

\title{CO Spectral Line Energy Distributions in Galactic Sources: Empirical Interpretation of Extragalactic Observations\footnote{{\it Herschel} is an ESA space observatory with science instruments provided by European-led Principal Investigator consortia and with important participation from NASA.}}
\shorttitle{CO SLEDs in Galactic Sources}
\shortauthors{Indriolo et al.}

\author{Nick Indriolo\altaffilmark{1} \& E. A. Bergin}
\affil{Department of Astronomy, University of Michigan, 1085 S. University Ave., Ann Arbor, MI 48109, USA}
\and
\author{J. R. Goicoechea \& J. Cernicharo}
\affil{Grupo de Astrof\'{i}sica Molecular, Instituto de Ciencia de Materiales de Madrid (CSIC). E-28049. Madrid, Spain}
\and
\author{M. Gerin \& A. Gusdorf}
\affil{LERMA, Observatoire de Paris, PSL Research University, CNRS, Sorbonne Universit\'{e}s, UPMC Univ. Paris 06, \'{E}cole normale sup\'{e}rieure, F-75005, Paris, France}
\author{D.~C.~Lis\altaffilmark{2}}
\affil{LERMA, Observatoire de Paris, PSL Research University, CNRS, Sorbonne Universit\'{e}s, UPMC Univ. Paris 06, F-75014, Paris, France}
\and
\author{P. Schilke}
\affil{I. Physikalisches Institut der Universit\"{a}t zu K\"{o}ln, Z\"{u}lpicher Str. 77, 50937 K\"{o}ln, Germany}
\altaffiltext{1}{Current address: Space Telescope Science Institute, Baltimore, MD 21218, USA; nindriolo@stsci.edu}
\altaffiltext{2}{California Institute of Technology, Cahill Center for Astronomy and Astrophysics 301-17, Pasadena, CA 91125, USA}

\begin{abstract}
The relative populations in rotational transitions of CO can be useful for inferring gas conditions and excitation mechanisms at work in the interstellar medium.  We present CO emission lines from rotational transitions observed with {\it Herschel}/HIFI in the star-forming cores Orion~S, Orion~KL, Sgr~B2(M), and W49N. Integrated line fluxes from these observations are combined with those from {\it Herschel}/PACS observations of the same sources to construct CO spectral line energy distributions (SLEDs) from $5\leq J_u\leq48$. These CO SLEDs are compared to those reported in other galaxies, with the intention of empirically determining which mechanisms dominate excitation in such systems. We find that CO SLEDs in Galactic star-forming cores cannot be used to reproduce those observed in other galaxies, although the discrepancies arise primarily as a result of beam filling factors. The much larger regions sampled by the {\it Herschel} beams at distances of several Mpc contain significant amounts of cooler gas which dominate the extragalactic CO SLEDs, in contrast to observations of Galactic star-forming regions which are focused specifically on cores containing primarily hot molecular gas.
\end{abstract}

\section{Introduction} \label{sec_intro}  

The {\it Herschel} Space Observatory \citep{pilbratt2010} enabled the first surveys of rotational transitions of CO with $4\leq J_{u}\leq50$ in emission throughout a wide sample of galaxies. These CO emission lines can be used to place constraints on the physical conditions (e.g., density, temperature, radiation field) within the emitting gas as the relative populations in the various rotational states are controlled by collisional and radiative (de)-excitation. The shape of the CO Spectral Line Energy Distribution (SLED)---flux in each emission line as a function of upper state energy---provides information about the gas conditions, and potentially the agent (e.g., shocks, X-rays, UV photons, cosmic rays) primarily responsible for heating the gas. Multiple observing programs targeted CO emission lines in the central regions of different types of galaxies---e.g., Ultra-Luminous InfraRed Galaxies (ULIRGs), Seyfert galaxies, starburst galaxies---for the purpose of determining which of the aforementioned mechanisms dominate the gas heating in each case. However, in most galaxies the observed CO SLEDs can be fit with a variety of models, such that it is difficult to conclude whether shocks, PDRs (photon dominated regions), or XDRs (X-ray dominated regions), are driving the CO excitation \citep[e.g,][]{hailey-dunsheath2012,kamenetzky2014,mashian2015,rosenberg2015}. While kinematic information would provide a clue to this puzzle, both the SPIRE \citep[Spectral and Photometric Imaging Receiver;][]{griffin2010SPIRE} and PACS \citep[Photoconductor Array Camera and Spectrometer;][]{poglitsch2010PACS} instruments used for these extragalactic observations are incapable of spectrally resolving the CO emission lines.

{\it Herschel} observations of CO emission lines have also been reported for a variety of regions within our Galaxy, including the well-studied objects Sgr~B2 \citep{etxaluze2013}, Sgr~A \citep{goicoechea2013}, and Orion~KL \citep{goicoechea2015}. In some Galactic sources, however, in addition to the low spectral resolution PACS data we also have {\it Herschel} HIFI \citep[Heterodyne Instrument for the Far Infrared;][]{degraauw2010} observations of $J_u\leq16$ transitions of CO that are spectrally resolved. By studying the velocity profiles of these CO emission lines in Galactic sources we can better constrain the excitation mechanisms involved in producing different CO SLED shapes.

Galactic regions that we consider herein include the Orion Bar, Orion South, Orion~KL, Sgr~B2(M), and W49N. The Orion Bar is a prototypical strongly illuminated PDR, and is located in the Orion star forming region at a distance of 414~pc \citep{menten2007}. It has the distinction of being nearly ``edge-on'', such that the atomic and molecular emission from different stratified layers can be studied \citep[e.g.,][]{vanderwiel2009,nagy2013}. Orion~S is an embedded star-forming region that contains multiple outflows \citep{ziurys1990,schmid-burgk1990,zapata2005,zapata2006}, shocked gas \citep{henney2007,rivilla2013}, and a PDR illuminated by the Trapezium stars \citep{peng2012}. In Section \ref{sec_analysis_hifi} we show that the CO emission profiles in Orion~S can be decomposed into these three components. Orion~KL is a luminous, high-mass star forming region comprised of multiple spatial and kinematic components. It harbors warm gas clumps, multiple suspected protostars, quiescent gas, and shocked gas resulting from explosive outflows \citep{blake1987,zapata2009,zapata2011,nissen2012,crockett2014}. Sgr~B2(M) is a compact, massive, star-forming core within the central molecular zone of our Galaxy, and its surrounding envelope is an X-ray reflection nebula \citep{murakami2000}. W49N is one of the most luminous and massive star-forming regions within our Galaxy \citep[][and references therein]{galvan-madrid2013}, harboring multiple star clusters and tens of O-type stars \citep{alves2003,wu2016}. It is frequently referred to as a starburst region in comparison to the eponymous class of galaxies. Through analysis of the CO SLEDs in these different well-studied, well-characterized Galactic regions, we aim to empirically interpret extragalactic CO SLEDs.

\section{Observations} \label{sec_obs}

The HEXOS \citep[{\it Herschel} observations of EXtra-Ordinary Sources; KPGT\_ebergin\_1;][]{bergin2010} key program includes scans over the full spectral range of HIFI (480--1906~GHz, with gaps from 1280--1430~GHz and 1540--1570~GHz) toward Orion~KL [($\alpha,\delta$)=($05^{\rm h}35^{\rm m}14^{\rm s}.3$, $-05\degr22\arcmin33.7\arcsec$)], Orion South [($\alpha,\delta$)=($05^{\rm h}35^{\rm m}13^{\rm s}.4$, $-05\degr24\arcmin08.1\arcsec$)], the Orion Bar [($\alpha,\delta$)=($05^{\rm h}35^{\rm m}20^{\rm s}.6$, $-05\degr25\arcmin14.0\arcsec$)], Sgr~B2(M) [($\alpha,\delta$)=($17^{\rm h}47^{\rm m}20^{\rm s}.35$, $-28\degr23\arcmin03.0\arcsec$)], and Sgr~B2(N) [($\alpha,\delta$)=($17^{\rm h}47^{\rm m}19^{\rm s}.88$, $-28\degr22\arcmin18.4\arcsec$)]. Targets were observed in dual beam switch (DBS) spectral scan mode with reference positions offset by 3\arcmin, and the wide band spectrometer (WBS) was employed to provide 1.1~MHz resolution. In all sources, the HIFI spectra cover rotational transitions of $^{12}$C$^{16}$O (hereafter referred to simply as CO) out of the $5\leq J_{u}\leq16$ levels.\footnote{The $J=12$--11 transition falls in the gap in frequency coverage.} Toward Orion~KL, however, the HIFI beam does not encompass the entire emitting region for $J_u\geq13$, so for these transitions there are separate pointings toward the hot core [($\alpha,\delta$)=(05:35:14.5, $-$05:22:30.9)] and compact ridge [($\alpha,\delta$)=(05:35:14.1, $-$05:22:36.5)] components \citep{crockett2014}. Individual analyses of the full spectra have been reported for Orion~KL \citep{crockett2014}, Orion~S \citep{tahani2016,tahani2013thesis}, the Orion Bar \citep{nagy2016arxiv}, and Sgr~B2(N) \citep{neill2014}. 

Unlike the full spectral scans made of Sgr~B2 and Orion sources, HIFI observations of W49N  [($\alpha, \delta$)=($19^{\rm h}10^{\rm m}13^{\rm s}.2$, $09\degr06\arcmin12.0\arcsec$)] were made in targeted spectral windows as part of multiple science and calibration programs. In total, 5 CO transitions in the HIFI frequency range were observed toward W49N. CO $J=6$--5 was covered by the PRISMAS (PRobing InterStellar Molecules with Absorption line Studies; KPGT\_mgerin\_1) key program in observations targeting D$_2$H$^+$ at 692~GHz, $J=7$--6 by OT1\_mgerin\_4 in observations targeting C~\textsc{i} at 809~GHz, and $J=8$--7 by OT1\_cvastel\_2 in observations targeting HDO at 919~GHz. Observations of these three transitions utilized the WBS, and were made in DBS mode. On the fly (OTF) maps of CO $J=10$--9 and $J=13$--12 were made as part of calibration observations. Table \ref{tbl_obsids} lists the ObsIDs that contain CO transitions for each source.

Sgr~B2, Orion KL, the Orion~Bar, and Orion~S were also observed as part of the HEXOS program with the PACS spectrometer over the $\sim$54--190~$\mu$m spectral range, covering the $14\leq J_{u}\leq48$ rotational transitions of CO.\footnote{Some wavelength ranges were affected by low spectral response (98-102\,$\mu$m) and spectral leakage (see the PACS observer's manual at \url{http://herschel.esac.esa.int/Docs/PACS/html/pacs_om.html}). We do not consider CO lines lying in these ranges.} PACS observations of the Orion~Bar will be presented by C. Joblin et al., (in preparation), so we do not discuss them further. The PRISMAS key program included PACS observations of W49N. The PACS spectrometer \citep{poglitsch2010PACS} provides 25 spectra over a $47\arcsec\times47\arcsec$ field of view resolved in 5$\times$5 spatial pixels (``spaxels''), each with an angular size of $9\farcs4\times9\farcs4$ on the sky. The resolving power of the grating spectrometer varies between $R\sim1000$--1500 ($\sim108$--190~$\mu$m range), $R\sim1700$--3000 (70--94~$\mu$m range) and $R\sim2700$--5500 (54--70~$\mu$m range). Spectra for all sources were obtained in the pointed Range Spectroscopy SED mode. Orion~S and W49N were observed in the standard ``chop-nod'' mode with a chopper throw of $\pm$6 arcmin. Owing to the very high far infrared continuum fluxes toward Sgr~B2(M) and Orion~KL (above the nominal saturation limits of PACS), these sources were observed in a specific non-standard engineering procedure \citep[{\it PacsCalWaveCalNo-ChopBurst}; see][]{goicoechea2015}. In order to avoid contamination from the bright Orion and Sgr~B2 extended clouds, the ``unchopped'' observing mode was used. In this mode, background subtraction is achieved by removing the telescope spectrum measured on a distant reference OFF-position (in our case separated by $\sim$20~arcmin).  ObsIDs corresponding to these data are also shown in Table \ref{tbl_obsids}. Herein, we focus our analysis solely on the CO rotational transitions within the PACS and HIFI spectral scans.

\section{Data Reduction}

\subsection{HIFI} \label{sec_hifiredux} 

As mentioned in Section \ref{sec_obs}, HIFI observations of CO in W49N are comprised of data from multiple programs, such that our data set is not uniform. Single pointing observations of the $J=6$--5 and $J=7$--6 transitions were processed to Level 2 using the standard HIPE \citep[{\it Herschel} Interactive Processing Environment;][]{ott2010} pipeline v12.0, and those of the $J=8$--7 transition were processed to Level 2 using HIPE v13.0. Baselines were subtracted using a first order polynomial and spectra taken at different LO frequencies were averaged together. Despite the beams for the H and V polarizations being separated by a few arcseconds, we found no evidence for sharp discrepancies between their associated spectra, so the two polarizations are also averaged together. OTF maps of the $J=10$--9 and $J=13$--12 transitions were processed to Level~2 via the standard HIPE pipeline v14.0 and converted to CLASS format. No spectrum in the OTF maps was taken at the same position as the single pointing observations. To extract a single spectrum at this position we take a weighted average of the spectra within the OTF maps. Individual spectra are weighted by $\exp(-r^2/\theta^2)$, where $r$ is the angular separation between each observation and the single pointing position and $\theta$ is the radius of the HIFI beam (i.e., gaussian half width at half maximum) at the transition frequency.

The full HIFI spectral scans of Orion and Sgr~B2 sources were processed using the same methods described in \citet{neill2014}. Orion~S was processed through HIPE v9.0 \citep{tahani2013thesis}, Orion~KL through HIPE v10.3, 
and Sgr~B2(M) through HIPE v8.0. These reduced data products are available for download via the {\it Herschel} Science Archive as user provided data products.\footnote{http://www.cosmos.esa.int/web/herschel/user-provided-data-products} All spectra were rescaled to account for the updated main beam efficiencies reported in Mueller et al. (2014).\footnote{The HIFI Beam: Release \#1; http://herschel.esac.esa.int/twiki/pub/Public/HifiCalibrationWeb/HifiBeamReleaseNote\_Sep2014.pdf Table 2 and Equation 8, therein.} Resulting CO emission lines are shown in Figure \ref{fig_cospectra}.

\subsection{PACS} \label{sec_pacsredux} 

PACS data were also processed using HIPE. Pointed observations with the PACS array do not provide fully spatially sampled maps. In particular, the individual spaxels do not fill the spectrometer point spread function (PSF) entirely. The measured width of the PSF is relatively constant for $\lambda\lesssim100$~$\mu$m (about the spaxel angular size) but increases at longer wavelengths. About 74\%\ (41\%) of the emission from a point source would fall in a given spaxel at  about 54~$\mu$m (190~$\mu$m). For sources with semi-extended emission this means that accurate line fluxes can only be extracted by adding the fluxes measured in apertures that cover several spaxels. Orion~KL and Sgr~B2 data were calibrated and reduced as described in \citet{goicoechea2015}. Reductions of Orion~S and W49N observations follow a similar method to that described in \citet{gerin2015}.

\section{Analysis}

\subsection{PACS Data}

After data reduction and line identification, a polynomial baseline was subtracted  in a narrow wavelength window around each detected CO line (with $J_u\geq14$). Line fluxes (in W\,m$^{-2}$) were extracted by fitting Gaussians to every line detected in every spaxel. Total line fluxes (within a given aperture) were obtained by summing the line fluxes measured in the different individual spaxels. In most cases we added the fluxes from all 25 spaxels within the $5\times5$ PACS array. However, fluxes in Orion~KL were determined as described in \citet{goicoechea2015} for a $3\times3$ spaxel ($\approx30\arcsec\times30\arcsec$) aperture centered on the hot core. This smaller region was used for extracting fluxes instead of the full array since the PACS footprint centered on Orion~KL is contaminated by Orion~H$_2$~Peak~1, which gives rise to most of the high-$J$ CO emission \citep[see Figure 1 in][]{goicoechea2015}. The line fluxes towards Sgr~B2(M) and Orion~S were extracted from the full $5\times5$ spaxel aperture. In W49N we extracted CO line fluxes both from the central spaxel alone with a point source correction applied, and from the full 5$\times$5 array. Our reasons for testing both methods are discussed in Section \ref{section_combPACSHIFI}. The PACS flux calibration accuracy is limited by detector response drifts and slight pointing offsets, and the absolute flux calibration accuracy is estimated to be on the order of 30\%.\footnote{\textit{PACS Spectroscopy performance and calibration}, PACS/ICC document ID PICC-KL-TN-041 (Vandenbussche et al.).} CO line fluxes determined from our analysis are reported in Table \ref{tbl_fluxes}.

\subsection{HIFI Data} \label{sec_analysis_hifi}

Our analysis of the CO $5\leq J_u\leq 16$ emission lines is performed using the spectra generated from the reduction described in Section \ref{sec_hifiredux}. Integrated line fluxes ($\int T_{MB}dv$) in units of K~km~s$^{-1}$ are extracted from the spectra in Figure \ref{fig_cospectra} as described below, and are converted to intensity and flux via
\begin{equation}
\mathrm{\frac{Intensity}{(W~m^{-2}~sr^{-1})}}=\frac{\nu^{3}2k_{b}}{c^{3}}\int T_{MB}dv=1.0248\times10^{-18}\left(\frac{\nu}{\mathrm{GHz} }\right)^{3}\left(\frac{\int T_{MB}dv}{\mathrm{K~km~s^{-1}}}\right)
\label{eq_Tdv_intensity}
\end{equation}
and
\begin{equation}
\mathrm{\frac{Flux}{(W~m^{-2})}}=\frac{\nu^{3}2k_{b}\Omega}{c^{3}}\int T_{MB}dv=1.0248\times10^{-18}\left(\frac{\Omega}{\mathrm{sr}}\right)\left(\frac{\nu}{\mathrm{GHz} }\right)^{3}\left(\frac{\int T_{MB}dv}{\mathrm{K~km~s^{-1}}}\right)
\label{eq_Tdv_flux}
\end{equation}
where $\nu$ is the transition frequency, $k_b$ is the Boltzmann constant, $c$ is the speed of light, and $\Omega$ is the main beam solid angle. Line fluxes are reported in Table \ref{tbl_fluxes}, along with $\Omega$ at the pertinent frequencies. 

\subsubsection{W49N}

CO emission lines in W49N are shown in the top left panel of Figure \ref{fig_cospectra}. They display the same double peaked profiles as observed in HCN and HCO$^+$ which are caused by strong self absorption and interpreted as a signature of gas infall \citep{roberts2011}. The self absorption scenario is favored over two separate emission components because the weaker isotopologue emission (e.g., H$^{13}$CO$^+$, H$^{13}$CN) is singly peaked at about 7~km~s$^{-1}$, the systemic velocity of W49N. We determine integrated line fluxes for CO over the velocity interval from $-40$ to 45 km~s$^{-1}$.

\subsubsection{Orion~KL}

The top right panel of Figure \ref{fig_cospectra} displays the CO emission lines observed toward Orion~KL. The broad line profiles likely contain emission from the various well-known components such as the hot core, compact ridge, and plateau (high velocity and low velocity outflows), with the extended ridge giving rise to some of the self absorption near 10~km~s$^{-1}$ \citep{blake1987,crockett2014}. Integrated line fluxes were determined over the velocity interval from $-$80 to 90 km~s$^{-1}$. Reported uncertainties account for the RMS noise level in the spectra (measured from $-125$ to $-100$~km~s$^{-1}$) and the weak emission features caused by other species that are within the velocity interval over which we measure integrated fluxes. Note that all of the weak emission features seen in the Orion~KL spectra at various LSR velocities are caused by molecules other than CO (e.g., SO$_2$ and CH$_3$OH). The emitting species have been identified and modeled by \citet{crockett2014}, and like the aforementioned reduced spectral scans, those models are also available as user provided data products.


\subsubsection{Sgr~B2(M)}

Toward Sgr~B2(M), CO emission is detected from all transitions covered by HIFI.  Self-absorption caused by the Sgr~B2 envelope is strong for transitions with $J_{u}\leq8$, moderately strong for transitions with $9\leq J_{u}\leq 11$, and either weak or not present for $J_{u}\geq13$ transitions.  Absorption due to gas in the foreground spiral arms is also seen for $J_{u}\leq8$ transitions at velocities blue-shifted from 63~km~s$^{-1}$, the systemic velocity of Sgr~B2(M). All of these features are shown in the bottom left panel of Figure \ref{fig_cospectra}. These results parallel those found for Sgr~B2(N) \citep{neill2014}. CO line fluxes reported in Table \ref{tbl_fluxes} for Sgr~B2(M) are integrated over the velocity range 0--130~km~s$^{-1}$.


\subsubsection{Orion S} \label{analysis_oris}

\begin{figure}
\epsscale{1.25}
\plotone{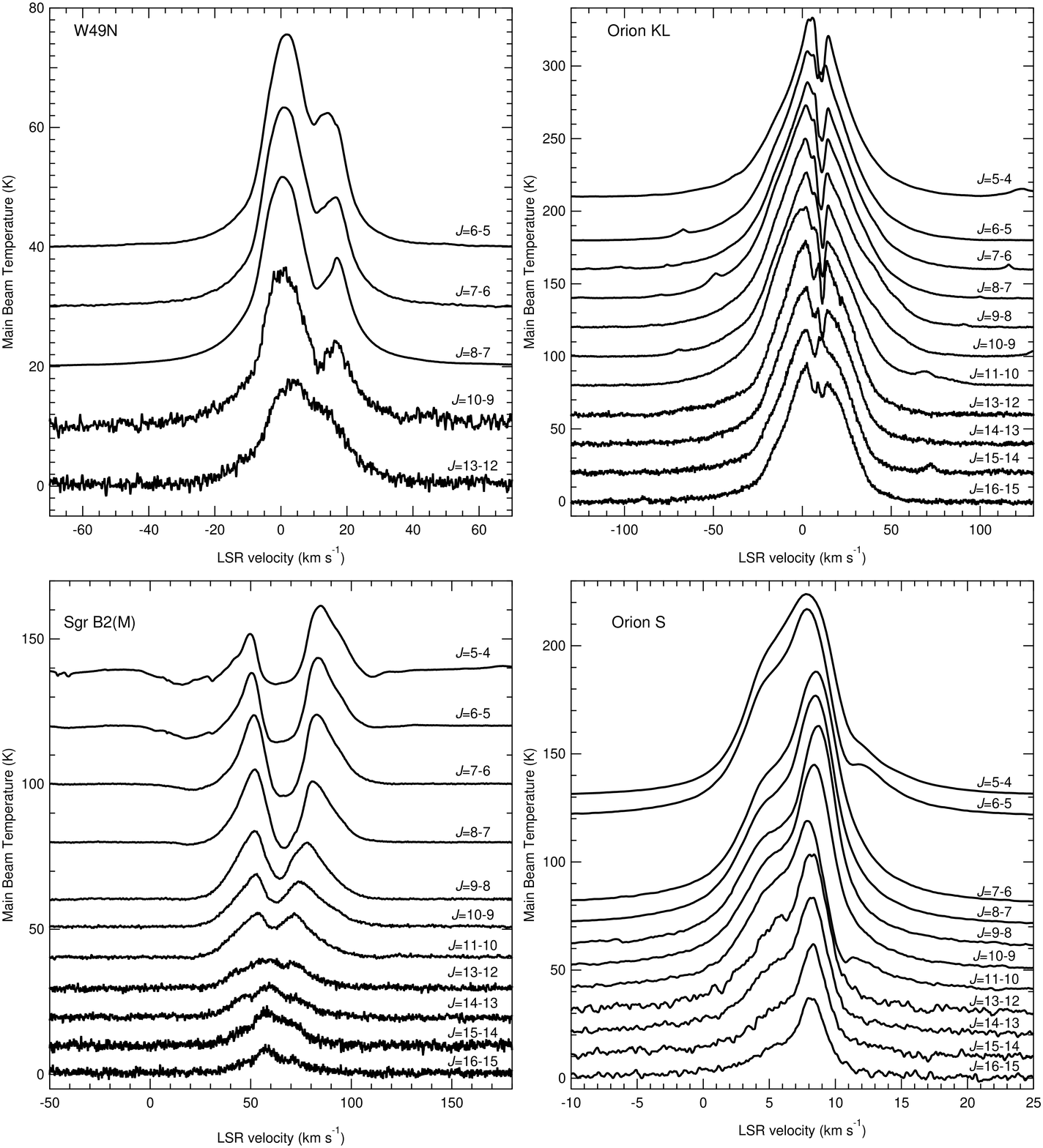}
\caption{CO emission lines in W49N (top left), Orion~KL (top right), Sgr~B2(M) (bottom left) and Orion S (bottom right). Note the different velocity and temperature scales in each panel. Spectra are shifted vertically for clarity. In Orion~KL the spectra displayed for the $13\leq J_u\leq16$ transitions are from the observations targeting the compact ridge.}
\label{fig_cospectra}
\end{figure}

\begin{figure}
\epsscale{1.25}
\plotone{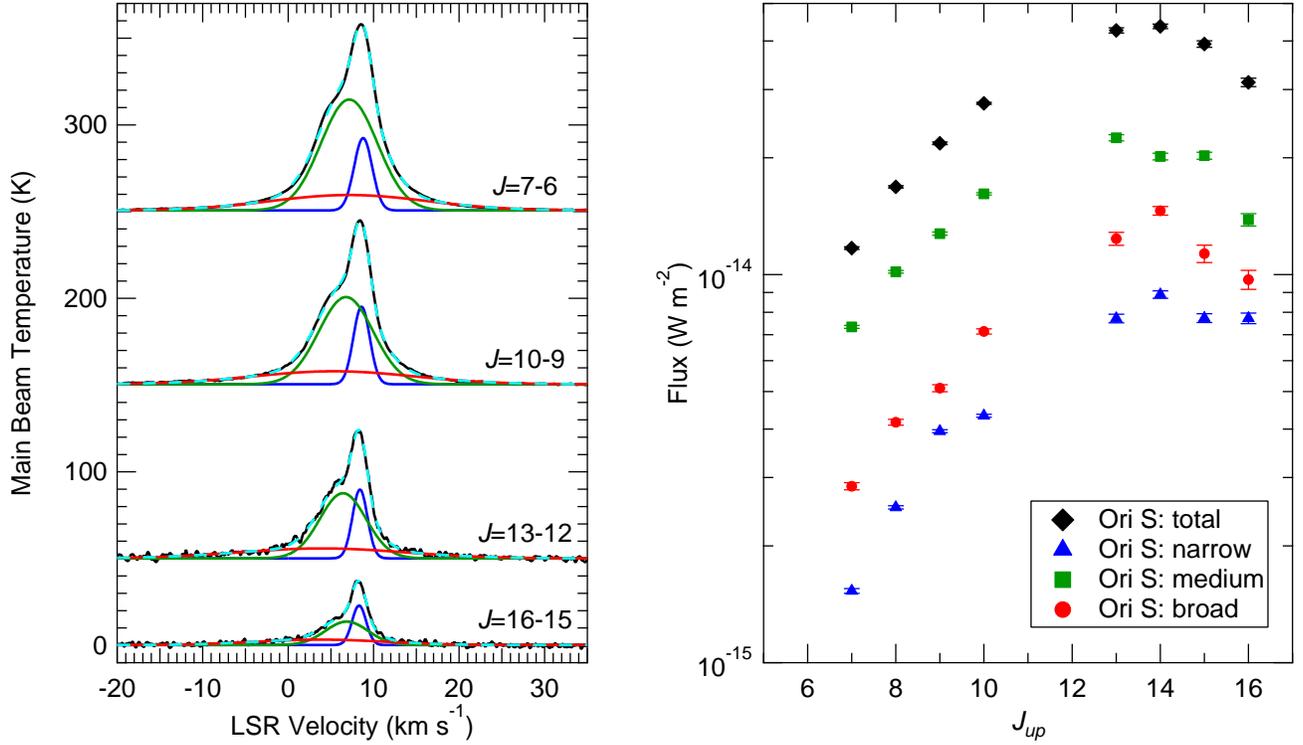}
\caption{The decomposition of select CO transitions toward Orion S into three gaussian components is displayed on the left. Spectra are shown in black, with the full fit given by dashed cyan curves. The narrow PDR component is shown in blue, the shock component in green, and the broad outflow component in red. Fluxes for each of the components---scaled as described in Section \ref{section_combPACSHIFI}---are plotted on the right side. Color coding matches the fit components in the left panel, although the total flux is marked by black (rather than cyan) diamonds. Flux uncertainties are generally smaller than the plotted markers.}
\label{fig_oris_decomp}
\end{figure}

\begin{figure}
\epsscale{0.8}
\plotone{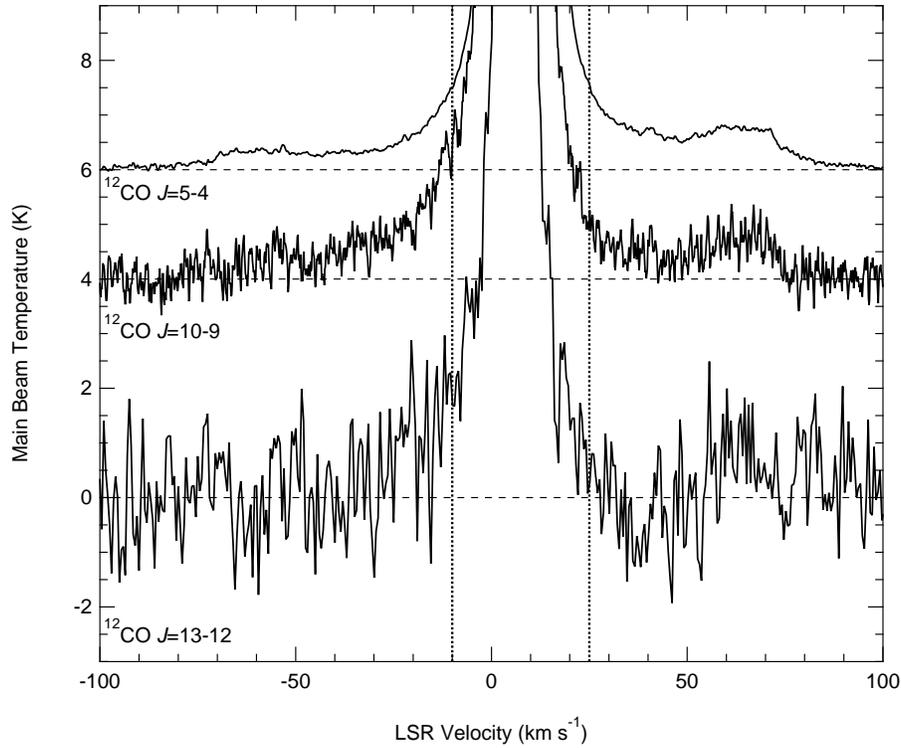}
\caption{Zoom in on select CO transitions toward Ori~S showing CO ``bullets''. Spectra have been shifted vertically for clarity, and the zero level for each spectrum is marked by a horizontal dashed line. The $J=13$--12 spectrum has been smoothed to 0.5~km~s$^{-1}$ resolution. Vertical dotted lines denote the full velocity range shown in the bottom right panel of Figure \ref{fig_cospectra}.}
\label{fig_oris_bullets}
\end{figure}

Orion~S is comprised of multiple physical components that can be distinguished via their differing kinematics. CO emission lines in Orion~S are shown in the bottom right panel of Figure \ref{fig_cospectra}, and they display a narrow (FWHM $\approx2$~km~s$^{-1}$) component centered at about 8.5~km~s$^{-1}$, a medium (FWHM $\approx7$~km~s$^{-1}$) component centered at about 7~km~s$^{-1}$, and a broad (FWHM $\approx20$~km~s$^{-1}$) component centered at about 5~km~s$^{-1}$. This structure mimics that of H$_2$O emission observed toward both low and high mass protostars \citep[e.g.,][]{kristensen2012,sanjosegarcia2016}. We fit these three components with gaussian functions, and our decomposition of the CO emission for select transitions is shown in left-hand side of Figure \ref{fig_oris_decomp}. The profile of the narrow component at 8.5~km~s$^{-1}$ is similar to that of CO emission in the Orion~Bar \citep{nagy2013}, and likely arises from the portion of Orion~S that is being illuminated by the Trapezium stars, i.e., in a PDR. The medium component accounts for most of the line flux and is likely due to shocks associated with protostellar activity within the region. The broad component likely corresponds to the outflows detected in SiO \citep{ziurys1990} and CO \citep{schmid-burgk1990}. A fourth component is visible in Figure \ref{fig_oris_bullets} as weak emission ($<1$~K) extending to high velocities ($\pm80$~km~s$^{-1}$), and is detected for all transitions with $J_{u}\leq11$. This emission corresponds to the highly collimated outflows observed in CO $J=2$--1 by \citet{zapata2005}. It is not detected in the $J_{u}\geq13$ transitions simply because the HIFI beam at these frequencies no longer encompasses any portion of the collimated outflows.

Careful inspection of the CO emission lines from Orion S in Figure \ref{fig_cospectra} reveals that the $J=5$--4, 6--5, and 11-10 profiles peak at velocities about 0.5~km~s$^{-1}$ lower than the other transitions, and show what appear as small absorption features at about 11~km~s$^{-1}$ . These artifacts are due to imperfect removal of CO emission that was in the reference beam at the off position \citep{tahani2016}. As a result, the fit parameters for the different gaussian components (i.e., $v_{\rm LSR}$, FWHM, and $\int Tdv$) found for these transitions do not agree with those found for the unaffected transitions, and we exclude them from the remainder of our analysis. Using our fits to the CO emission lines we can generate SLEDs for each of the three different components as shown in the right-hand side of Figure \ref{fig_oris_decomp}. All three components show roughly the same overall shape, with SLEDs peaking around $J_{u}=14$.

\subsection{Combining PACS and HIFI Results} \label{section_combPACSHIFI}

For Orion~S, Orion~KL, and Sgr~B2(M) we have determined fluxes for the $J_u=14$, 15, and 16 transitions from both HIFI and PACS observations. As can be seen in Table \ref{tbl_fluxes}, the values determined from the two different instruments do not agree. To generate a CO SLED without discontinuities, we scale HIFI fluxes given the following reasoning. In the limit of an unresolved point source, the flux measured is independent of the beam size. In the limit of a resolved source with uniform emission, the intensity measured is independent of beam size. By comparing PACS fluxes to HIFI fluxes and PACS intensities to HIFI intensities, we can determine which scenario is more likely applicable for each source. For Orion~S, the PACS fluxes reported in Table \ref{tbl_fluxes} come from the full 5$\times$5 spaxel array (47$\times$47~arcsec). Fluxes for the $J_u=14$, 15, and 16 transitions determined from HIFI are lower than those determined from PACS, while intensities determined from HIFI are higher than those determined from PACS. For all three transitions, a roughly constant scaling factor (6\% variation) can be used to convert PACS intensities to HIFI intensities, but not to convert PACS fluxes to HIFI fluxes ($\sim50$\% variation). This potentially indicates a source size that is larger than the HIFI beam(s), but smaller than the PACS 5$\times$5 footprint, a scenario confirmed by inspection of the mid-$J$ CO emission lines observed in each individual spaxel toward Orion~S.
If so, then the fluxes measured by PACS are correct, while those measured by HIFI are too low. Similarly, the intensities measured by HIFI are correct, while those measured by PACS are underestimated. Dividing the PACS flux by the HIFI intensity gives the source size (assuming uniform emission), and this can be used to scale the HIFI fluxes to the values that would have been measured had the beams fully encompassed the emitting region. 
In this way, we remove the discrepancies between the PACS and HIFI fluxes reported in Table \ref{tbl_fluxes}, which would otherwise appear as discontinuities in the various CO SLEDs. The scaling factors used in this conversion for each source are: $2.58\times10^{-8}/\Omega$ for Orion S, $1.11\times10^{-8}/\Omega$ for Orion~KL, and $9.21\times10^{-9}/\Omega$ for Sgr~B2(M), where $\Omega$ is the beam solid angle at the transition frequency as reported in Table \ref{tbl_fluxes}.

\begin{figure}
\epsscale{1.1}
\plotone{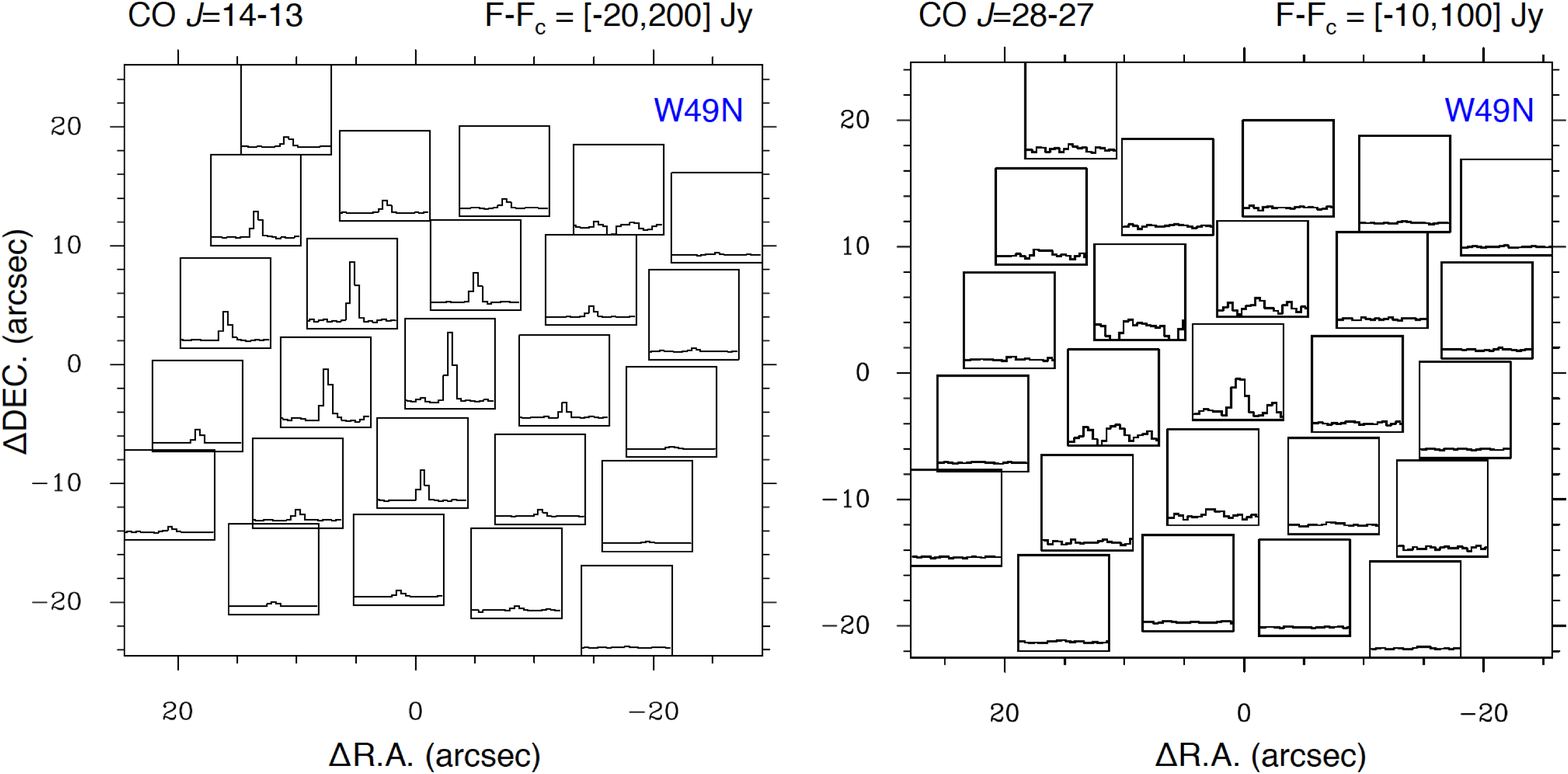}
\caption{Continuum-subtracted high-$J$ CO maps for W49N obtained with the PACS array in 25 spaxels. The line flux scale (in Jy) is indicated in the top-right of each panel. The $(0,0)$ position corresponds to ($\alpha, \delta$)=($19^{\rm h}10^{\rm m}13^{\rm s}.1$, $09\degr06\arcmin12.0\arcsec$).}
\label{fig_w49n_map}
\end{figure}

For W49N no CO transition was observed by both PACS and HIFI. Examination of CO spectra observed with each PACS spaxel reveals that the emitting region changes as a function of upper state energy. The lowest lines show emission over multiple spaxels, while the highest lines are concentrated in only the central spaxel. This is demonstrated in Figure \ref{fig_w49n_map} where the $J$=14--13 and $J=28$--27 spectra in each PACS spaxel show differences in the emitting region. Because of the changing source size, CO line fluxes extracted from the full 5$\times$5 array are larger than those extracted from the central spaxel alone with a point source correction applied, except at high $J_u$ where the emission becomes concentrated. The HIFI beams for the observed CO transitions are larger than 1 spaxel, smaller than the full PACS array, and also smaller than the CO emitting region seen in the HIFI OTF integrated intensity maps, further complicating the analysis. We choose to use the 5$\times$5 PACS fluxes throughout the remainder of our analysis in order to avoid ``throwing away'' flux from the mid-$J$ CO lines. HIFI fluxes are scaled by $1.23\times10^{-8}/\Omega$ to remove the discontinuity between PACS and HIFI fluxes. All of these issues highlight the difficulties inherent in combining emission line fluxes extracted from detectors with different beam sizes covering different portions of a target region that itself changes size and shape as a function of transition energy.

\section{Discussion}

\begin{figure}
\epsscale{0.9}
\plotone{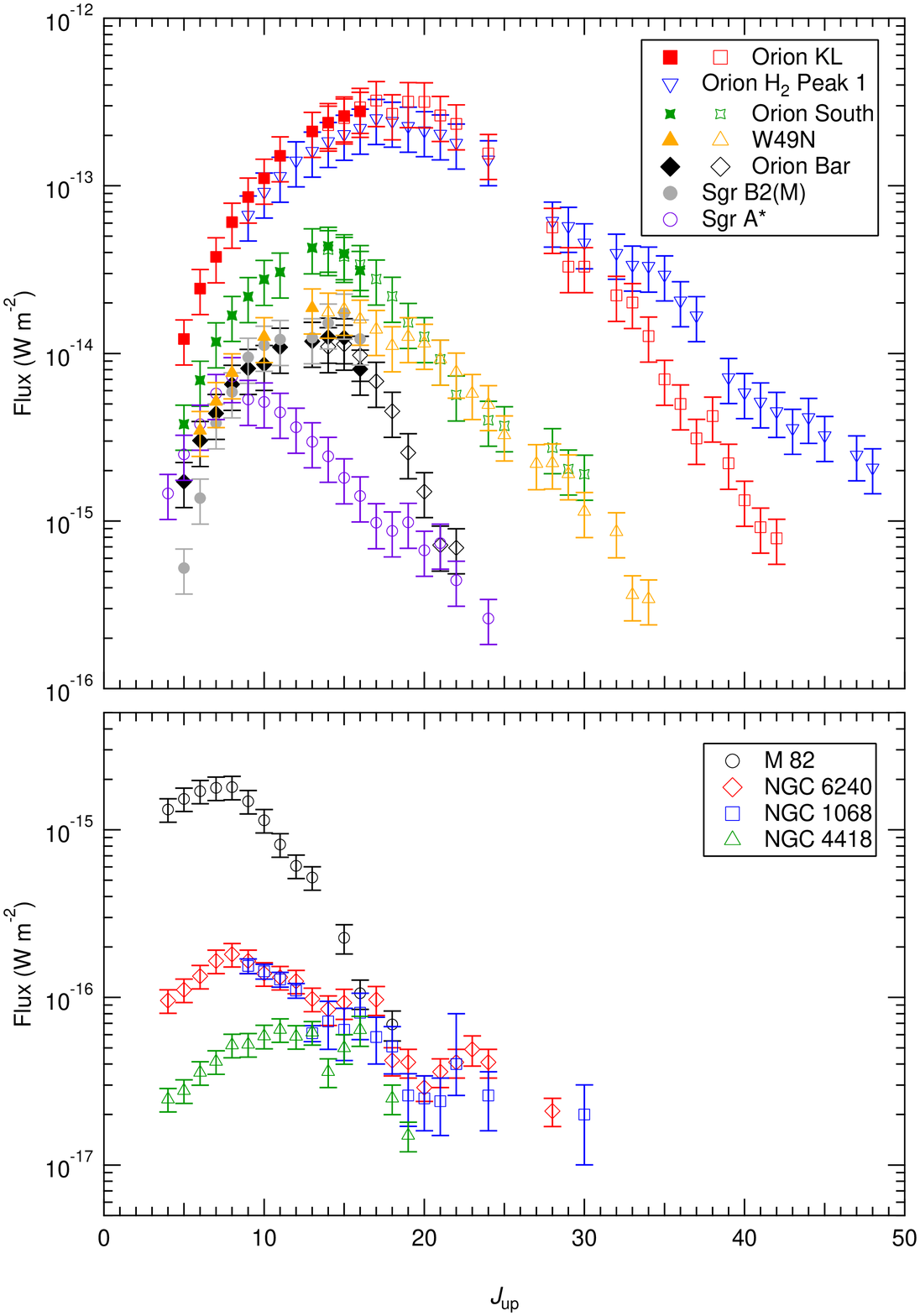}
\caption{CO spectral line energy distributions for Galactic sources (top), and for extragalactic sources (bottom). Filled symbols denote fluxes determined from HIFI observations, and open symbols from either SPIRE ($J_u\leq13$) or PACS ($J_u\geq14$) observations. The 30\%\ uncertainties in PACS fluxes have been applied to the HIFI fluxes as well given our scaling procedure. References for published line fluxes are as follows: Orion~H$_2$ Peak 1 \citep{goicoechea2015}; Orion Bar (C. Joblin in prep).; Sgr~A* \citep{goicoechea2013};  NGC~1068 \citep{spinoglio2012,hailey-dunsheath2012,janssen2015}; NGC 6240 \citep{rosenberg2015,mashian2015}; M 82 \citep{kamenetzky2012,mashian2015}; NGC 4418  \citep{rosenberg2015,mashian2015}.}
\label{fig_cosleds}
\end{figure}

The CO SLEDs resulting from the analysis described above are shown in the top panel of Figure \ref{fig_cosleds} for Sgr~B2(M), Orion~KL, Orion~S, and W49N. Additionally, the top panel displays CO SLEDs in the Galactic sources Orion~H$_2$~Peak~1 \citep[][]{goicoechea2015}, the Orion Bar (C. Joblin in preparation), and Sgr~A* \citep[][]{goicoechea2013}, while the bottom panel shows CO SLEDs from the Seyfert~2 galaxy NGC~1068 \citep[][]{spinoglio2012,hailey-dunsheath2012,janssen2015}, luminous infrared galaxies NGC~6240 \citep[][]{mashian2015,rosenberg2015} and NGC~4418 \citep[][]{mashian2015,rosenberg2015}, and the starburst galaxy M~82 \citep[][]{kamenetzky2012,mashian2015}. The vertical axes in both the top and bottom panels have been scaled to facilitate direct visual comparison of the CO SLED shapes. We focus first on the CO SLEDs in Galactic sources.

All of the Orion sources are at a distance of about 420~pc \citep{menten2007}, so differences in those SLEDs are mostly intrinsic to the sources. The Orion Bar shows the simplest profile, and can be considered a template for the CO SLED in a strongly illuminated PDR with $\chi_{\rm UV}\approx10^{4}$ \citep[expressed in units of the mean interstellar radiation field from][]{draine1978}. The other Orion sources likely contain PDR components as well, as some portion of the gas is being illuminated by FUV photons from the Trapezium cluster. Indeed, the decomposition of Orion~S line profiles described in Section \ref{analysis_oris} shows this PDR component, and demonstrates that it has the smallest contribution to the total emission line flux. Orion~S and Orion~KL are both regions of embedded massive star formation, and the internal energy provided by this process through outflows, shocks, and radiation serves to increase the CO line flux compared to the externally heated Orion Bar. Orion~KL has a bolometric luminosity about 10 times that of Orion~S \citep[][and references therein]{odell2008}, hence the larger CO line fluxes. Simply put, the increasing energy available going from the Orion Bar to Orion~S to Orion~KL both excites a larger amount of molecular gas and pushes population in the rotational levels of CO to higher $J$, thus producing the observed CO SLEDs. The SLED for Orion~H$_2$ Peak~1---a region which can be considered a prototypical strong molecular shock---has a different shape with line fluxes decreasing more slowly as $J_u$ increases. Excitation in this region is dominated by shock heating as a high-velocity outflow collides with quiescent molecular gas, and emission from the highest $J_u$ transitions arises from hot ($T\sim3000$~K), dense ($n\sim10^7$~cm$^{-3}$) gas \citep{goicoechea2015}.

The Sgr~B2(M) and Orion Bar SLEDs are very similar, despite the sources themselves and their CO emission line profiles being vastly different. As shown in Figure \ref{fig_cospectra} the Sgr~B2(M) line profiles are complex and dominated by self-absorption from the envelope that surrounds the hot core, whereas the Orion Bar has a single velocity component in emission \citep[see, e.g., Figure 2 in][]{nagy2013}. Additionally, far-infrared extinction by the Sgr~B2 envelope may reduce the observed flux in the $J=16$--15 transition by a factor of 10 below that actually produced by Sgr~B2(M) \citep{etxaluze2013}, and by even larger factors for higher $J_u$ lines (hence the non-detections for $J_u>16$ transitions). While far IR extinction only affects certain objects, it is clear that attempting to infer source attributes from the CO line fluxes alone is a complicated and likely degenerate procedure.

The CO SLED in W49N most closely resembles that in Orion~S. W49N is one of the most luminous massive star forming regions in our Galaxy---about $10^3$ times more luminous than Orion~S \citep{sievers1991}---and has been considered a template for extragalactic giant H~\textsc{ii} regions \citep{wu2016}, so this similarity is unexpected. \citet{nagy2012} concluded that both UV photons and mechanical processes (e.g., winds and outflows) are likely the dominant heating mechanisms in this region, while X-rays do not contribute significantly. It seems likely then that Orion~S experiences similar conditions, just on a much smaller scale. Of the Galactic sources shown in Figure \ref{fig_cosleds}, the CO SLED in Sgr~A* peaks at the lowest value of $J_u$. \citet{goicoechea2013} concluded that UV photons and shocks are responsible for heating the hot molecular gas giving rise to the CO emission near Sgr~A*, and that presently neither X-rays nor cosmic rays play a large role.

A multitude of galaxies have been observed by {\it Herschel} with SPIRE \citep[e.g.,][]{kamenetzky2014,rosenberg2015} and PACS \citep[e.g.,][]{mashian2015} with CO emission lines as a primary target. We selected the 4 galaxies shown in the bottom panel of Figure \ref{fig_cosleds} for comparison to our Galactic sources because they present a variety of CO SLED shapes, including a member of each of the three classes defined by \citet{rosenberg2015}. Additionally, NGC~1068 and NGC~6240 have CO emission detected out to the highest $J_u$ of any galaxies, providing the most extensive comparisons to Galactic regions, and the CO SLED of NGC~4418 peaks at the highest $J_u$ of any galaxy. Note that we have excluded the SPIRE observations of NGC~1068 for $4\leq J_u\leq8$ as the larger beam at these frequencies contains two emission regions---circumnuclear disk (CND) and extended ring---whereas the SPIRE $J_u\geq9$ and PACS $J_u\geq14$ observations only probe the nuclear disk \citep{spinoglio2012,hailey-dunsheath2012}. 

It is immediately apparent that the extragalactic CO SLEDs differ in shape from their Galactic counterparts. In fact, linear combinations of the Orion, W49N, and Sgr~B2(M) CO SLEDs are incapable of reproducing those seen in M~82, NGC~1068, and NGC~6240 because all of these Galactic SLEDs peak at higher $J_u$ than the extragalactic SLEDs. Even the individual components of Orion~S shown in Figure \ref{fig_oris_decomp} fail in this regard. Only the Sgr~A* CO SLED peaks at low enough $J_u$ that it could conceivably be used to re-construct the extragalactic sources, while only the NGC~4418 CO SLED peaks at high enough $J_u$ that it could conceivably be reproduced by Galactic sources.  These disparities effectively prevent the empirical interpretation of extragalactic CO SLEDs and their underlying excitation mechanisms, yet simultaneously beg the question: Why do most Galactic CO SLEDs not resemble those in other galaxies?

The clearest difference between Galactic and extragalactic CO SLEDs is where the distribution peaks (i.e., which CO emission line has the largest flux). The NGC~4418 CO SLED peaks at $J_u\approx11$--13, the NGC~6240 and M~82 SLEDs at $J_u=8$, while the NGC 1068 SLED is increasing toward lower $J_u$, with $J_u=9$ being the lowest transition where the measured flux arises solely in the CND component. The Orion~KL and Orion~H$_2$~Peak~1 SLEDs peak near $J_u=18$ with emission in Peak~1 extending all the way to $J_u=48$, while the Orion~Bar, Orion~S, W49N, and Sgr~B2(M) SLEDs peak near $J_u=14$. Sgr~A* has a CO SLED that peaks at $J_u=8$. Where the CO SLEDs peak depends on the physical conditions of the gas, with hotter, denser gas leading to increased population, and thus flux, for higher $J_u$ transitions. While it may at first seem counterintuitive that Galactic star forming regions appear to harbor more hot, dense molecular gas than regions surrounding active galactic nuclei (AGN), this is likely an effect of beam filling factors. At a distance of 14.4~Mpc \citep{bland-hawthorn1997} the 9\farcs4$\times$9\farcs4 central spaxel of PACS covers a region about 650~pc on a side in NGC~1068. For NGC~4418 at $d=34$~Mpc \citep{braatz1997} and NGC~6240 at $d=107$~Mpc \citep{meijerink2013} the central spaxel covers a region approximately 1.5$\times$1.5~kpc and 4.6$\times$4.6~kpc, respectively. CO line fluxes in M~82 were extracted from the full 47\arcsec$\times$47\arcsec\ PACS array \citep{mashian2015}, which covers a region 770$\times$770~pc at $d=3.4$~Mpc \citep{dalcanton2009}. These regions, which do contain hot, dense gas as evidenced by emission from high-$J_u$ CO, must also contain large amounts of cooler gas which emit primarily at lower $J_u$ transitions of CO. Emission from the hot, dense gas only fills a small portion of the beam, while emission from the more extended cooler gas fills a much larger portion of the beam and ends up dominating the CO SLED.

Beam filling effects have previously been invoked to describe CO SLEDs observed in the Galactic center. One scenario proposed by \citet{goicoechea2013} to explain the Sgr~A* CO SLED suggests that the hot gas responsible for the high-$J_u$ emission is concentrated in small dense clumps that reside in a more diffuse, extended medium which gives rise to the lower $J_u$ emission. Another study by \citet{kamenetzky2014} compared the CO SLEDs of Sgr~B2(M), Sgr~B2(N), and the Sgr~B2 envelope determined from SPIRE observations \citep{etxaluze2013}, to those of several other galaxies. They note that the CO SLED of the Sgr~B2 envelope resembles those of other galaxies, while the CO SLEDs in the Sgr~B2 cores (i.e., those specifically focused on hot gas) peak at higher $J_u$. Furthermore, \citet{kamenetzky2014} conclude that while CO emission from star-forming cores must be present in their observations of other galaxies, the line flux is dominated by emission from warm, extended molecular clouds. In contrast, the observations of Galactic sources are tightly focused on known energetic regions. The full PACS footprint covers an area of about 0.1$\times$0.1~pc in the Orion star-forming region ($d\approx420$~pc), and about 2.5$\times$2.5~pc in W49N \citep[$d=11.1\pm0.8$~kpc;][]{zhang2013}. With hot, dense gas filling a large portion of the beam and a lack of ``contamination'' from unassociated cooler gas, the Galactic CO SLEDs peak at higher $J_u$. 


\section{Summary}

We have observed rotational transitions from the $5\leq J_u\leq16$ states of CO with HIFI and from $J_u\geq14$ with PACS in emission toward Orion~S, Orion~KL, Sgr~B2(M), and W49N. Fluxes are extracted from the CO emission lines and used to construct spectral line energy distributions (SLEDs). Our original intent was to empirically interpret CO SLEDs in other galaxies by reconstructing them from linear combinations of CO SLEDs in Galactic sources where the gas properties and heating mechanisms are well characterized. However, the CO SLEDs in our sample of Galactic sources all peak at higher $J_u$ than the CO SLEDs observed in other galaxies, such that no combination can successfully reproduce the extragalactic observations. The difference between Galactic and extragalactic CO SLEDs is primarily a beam filling effect. Our observations in the Milky Way specifically target star-forming cores, preferentially sampling hot molecular gas while excluding cold quiescent gas, such that the resulting CO SLEDs peak around $14\lesssim J_u\lesssim20$. In other galaxies the PACS and SPIRE beams cover much larger physical regions than they do within the Milky Way. As a result, in addition to the small pockets of hot, dense gas which produce high-$J_u$ CO emission, these beams also sample a large amount of cooler, more extended gas. It is this material, filling a much larger portion of the beam than the hot dense gas, which dominates the CO emission and causes SLEDs to peak closer to $J_u\sim8$. As such, we urge that careful consideration be given to these effects when comparing {\it Herschel} observations sampling vastly different physical size scales.

\mbox{}

The authors thank the anonymous referee for suggestions to improve the clarity of the paper. Support for this work was provided by NASA through an award issued by JPL/Caltech. J.R.G. and J.C. thank the ERC for support under grant ERC-2013-Syg-610256-NANOCOSMOS, and the Spanish MINECO for support under grant AYA2012-32032. HIFI has been designed and built by a consortium of institutes and university departments from across Europe, Canada and the United States under the leadership of SRON Netherlands Institute for Space Research, Groningen, The Netherlands and with major contributions from Germany, France and the US. Consortium members are: Canada: CSA, U.Waterloo; France: CESR, LAB, LERMA, IRAM; Germany: KOSMA, MPIfR, MPS; Ireland, NUI Maynooth; Italy: ASI, IFSI-INAF, Osservatorio Astrofisico di Arcetri-INAF; Netherlands: SRON, TUD; Poland: CAMK, CBK; Spain: Observatorio Astronómico Nacional (IGN), Centro de Astrobiología (CSIC-INTA). Sweden: Chalmers University of Technology - MC2, RSS \& GARD; Onsala Space Observatory; Swedish National Space Board, Stockholm University - Stockholm Observatory; Switzerland: ETH Zurich, FHNW; USA: Caltech, JPL, NHSC. 

PACS has been developed by a consortium of institutes led by MPE (Germany) and including UVIE (Austria); KU Leuven, CSL, IMEC (Belgium); CEA, LAM (France); MPIA (Germany); INAF-IFSI/OAA/OAP/OAT, LENS, SISSA (Italy); IAC (Spain). This development has been supported by the funding agencies BMVIT (Austria), ESA-PRODEX (Belgium), CEA/CNES (France), DLR (Germany), ASI/INAF (Italy), and CICYT/MCYT (Spain).

\bibliographystyle{aasjournal}
\bibliography{indy_master}


\clearpage
\begin{deluxetable}{crrrr}
\tabletypesize{\small}
\tablecaption{Observation Identifiers (ObsIDs) for Spectra Containing CO Emission\label{tbl_obsids}}
\tablehead{ & \colhead{Orion~S} & \colhead{Orion~KL} & \colhead{Sgr~B2(M)} & \colhead{W49N}
}
\startdata
 & \multicolumn{4}{c}{HIFI} \\
\hline
CO $J=5$--4 & 1342204001 & 1342191592 & 1342191565 & \\
CO $J=6$--5 & 04708 & 194540 & 192546 & 1342194554, 5, 6 \\
CO $J=7$--6 & 05332 & 205334 & 204723 & 1342230253, 4, 5 \\
CO $J=8$--7 & 05336 & 192329 & 206455 & 1342244816, 7, 8 \\
CO $J=9$--8 & 03150 & 191601 & 218200 &  \\
CO $J=10$--9 & 05871 & 191725 & 204739 & 1342253940, 68481 \\
CO $J=11$--10 & 16384 & 216387 & 215935 &  \\
CO $J=13$--12 & 03925 & 1342191727, 8 & 192656 & 1342254900, 68195 \\
CO $J=14$--13 & 03948 & 1342191562, 3 & 206501 &  \\
CO $J=15$--14 & 05534 & 1342194732, 3 & 216702 &  \\
CO $J=16$--15 & 05540 & 1342192673, 4 & 206640 & \\
\hline
 & \multicolumn{4}{c}{PACS} \\
\hline
CO $J=14$--13 through 35--34 & 1342218570 & 1342218575 & 1342206883 & 1342207774\\
CO $J=36$--35 through 48--47 & 1342218569 & 1342218576 &            &  1342207775 \\
\enddata
\end{deluxetable}
\normalsize
\clearpage


\begin{deluxetable}{lccccc}
\tabletypesize{\scriptsize}
\tablecaption{CO Line Fluxes\label{tbl_fluxes}}
\tablehead{
 & & \colhead{Orion~S} & \colhead{Orion~KL} & \colhead{Sgr~B2(M)} & \colhead{W49N} \\
 \hline
 & \colhead{$\Omega$} & \multicolumn{4}{c}{Flux} \\
\colhead{Transition} & \colhead{($10^{-9}$~sr)} & \colhead{($10^{-15}$ W~m$^{-2}$)} & \colhead{($10^{-15}$ W~m$^{-2}$)} & \colhead{($10^{-15}$ W~m$^{-2}$)} & \colhead{($10^{-15}$ W~m$^{-2}$)}
 }
 \startdata
HIFI & & & & & \\
   \hline
 $J=5$--4  &   24.0 &   3.53$\pm$0.05 &     26.2$\pm$0.3 &     1.37$\pm$0.03 &               ... \\
 $J=6$--5  &   17.1 &   4.59$\pm$0.06 &     37.5$\pm$0.4 &     2.54$\pm$0.01 &     4.80$\pm$0.24 \\
 $J=7$--6  &   12.3 &   5.61$\pm$0.04 &     41.8$\pm$0.5 &     5.16$\pm$0.02 &     5.16$\pm$0.26 \\
 $J=8$--7  &   9.44 &   6.17$\pm$0.04 &     51.5$\pm$0.5 &     6.05$\pm$0.16 &     5.86$\pm$0.29 \\
 $J=9$--8  &   7.52 &   6.37$\pm$0.05 &     57.9$\pm$0.6 &     7.74$\pm$0.11 &               ... \\
$J=10$--9  &   6.64 &   7.12$\pm$0.04 &     66.1$\pm$0.7 &     8.08$\pm$0.21 &     6.79$\pm$0.34 \\
$J=11$--10 &   5.49 &   6.51$\pm$0.05 &     74.3$\pm$0.8 &     7.23$\pm$0.16 &               ... \\
$J=12$--11 &    ... &             ... &              ... &               ... &               ... \\
$J=13$--12 &   3.59 &   5.94$\pm$0.09 &     67.9$\pm$0.9 &     4.82$\pm$0.14 &     5.43$\pm$0.27 \\
$J=14$--13 &   3.09 &   5.25$\pm$0.07 &     66.1$\pm$1.0 &     5.11$\pm$0.29 &               ... \\
$J=15$--14 &   2.78 &   4.25$\pm$0.08 &     65.4$\pm$1.1 &     5.25$\pm$0.10 &               ... \\
$J=16$--15 &   2.45 &   2.97$\pm$0.07 &     61.2$\pm$1.1 &     3.24$\pm$0.14 &               ... \\
   \hline
     PACS  &        &      $5\times5$ &       $3\times3$ &        $5\times5$ &        $5\times5$ \\
   \hline
$J=14$--13 &        &           41.58 &            230.3 &             18.00 &             17.60 \\
$J=15$--14 &        &           37.93 &            253.9 &             15.28 &             18.30 \\
$J=16$--15 &        &           33.89 &            294.2 &             11.41 &             16.00 \\
$J=17$--16 &        &           27.78 &            323.0 &                   &             13.90 \\
$J=18$--17 &        &           21.91 &            269.0 &                   &             11.10 \\
$J=19$--18 &        &           15.34 &            318.4 &                   &             12.60 \\
$J=20$--19 &        &           12.65 &            317.5 &                   &             11.50 \\
$J=21$--20 &        &            9.23 &            263.4 &                   &              9.23 \\
$J=22$--21 &        &            5.65 &            233.9 &                   &              7.73 \\
$J=23$--22 &        &           14.97 &              ... &                   &              5.76 \\
$J=24$--23 &        &            3.99 &            156.4 &                   &              4.95 \\
$J=25$--24 &        &            3.70 &              ... &                   &              3.26 \\
$J=26$--25 &        &             ... &              ... &                   &               ... \\
$J=27$--26 &        &             ... &              ... &                   &              2.20 \\
$J=28$--27 &        &            2.74 &             56.5 &                   &              2.22 \\
$J=29$--28 &        &            2.04 &             32.9 &                   &              1.91 \\
$J=30$--29 &        &            1.90 &             32.9 &                   &              1.14 \\
$J=31$--30 &        &                 &              ... &                   &               ... \\
$J=32$--31 &        &                 &             22.2 &                   &              0.86 \\
$J=33$--32 &        &                 &             20.1 &                   &              0.36 \\
$J=34$--33 &        &                 &             12.7 &                   &              0.34 \\
$J=35$--34 &        &                 &              7.0 &                   &                   \\
$J=36$--35 &        &                 &              5.0 &                   &                   \\
$J=37$--36 &        &                 &              3.1 &                   &                   \\
$J=38$--37 &        &                 &              4.2 &                   &                   \\
$J=39$--38 &        &                 &              2.2 &                   &                   \\
$J=40$--39 &        &                 &              1.3 &                   &                   \\
$J=41$--40 &        &                 &              0.9 &                   &                   \\
$J=42$--41 &        &                 &              0.8 &                   &                   \\
\enddata
\tablecomments{Absolute flux calibration accuracy is estimated to be about 30\%\ for PACS.}
\end{deluxetable}
\normalsize
\clearpage


\end{document}